\documentclass[rnote, traditabstract, usegraphicx]{aa}

\usepackage{natbib}
\bibpunct{(}{)}{;}{a}{}{,} 
\usepackage{subfigure}
\usepackage{subfig}
\usepackage{caption}
\usepackage{longtable}
\usepackage{graphicx}
\usepackage{graphics}
\usepackage{keyval}
\usepackage{trig}
\usepackage[american]{babel}
\usepackage{url}
\usepackage{color}
\usepackage{amsmath}
\usepackage{bm}
\usepackage{longtable}

\def\beq{\begin{equation}}
\def\eeq{\end{equation}}
\def\bey{\begin{eqnarray}}
\def\eey{\end{eqnarray}}

\def\lsim{\mathrel{\raise.3ex\hbox{$<$\kern-.75em\lower1ex\hbox{$\sim$}}}}
\def\gsim{\mathrel{\raise.3ex\hbox{$  $\kern-.75em\lower1ex\hbox{$\sim$}}}}

\def\a0{A_{\mathrm{0}}}

\def\apjl{ApJ}

\def\aap{A\&A}      
\def\apjs{ApJS}       

\newcommand\Fig[1]{Fig.~\ref{#1}}
\newcommand\Tab[1]{Table~\ref{#1}}
\newcommand\Sec[1]{Sec.~\ref{#1}}

\makeatletter\def\CT{\def\@captype{figure}}\makeatother

\begin{document}

   \title{RR Lyrae mode switching in globular cluster M 68 (NGC~4590)}

   \author{N. Kains\inst{\ref{stsci}}\and
          D. M. Bramich\inst{\ref{qeeri}} \and 
          A. Arellano Ferro\inst{\ref{unam}}\and
         R. Figuera Jaimes\inst{\ref{standrews}, \ref{eso}}
          }

\institute{Space Telescope Science Institute, 3700 San Martin Drive, Baltimore, MD 21218, United States of America \label{stsci}\\
\and Qatar Environment and Energy Research Institute, Qatar Foundation, P.O. Box 5825, Doha, Qatar \label{qeeri}\\
\and Instituto de Astronom\'{i}a, Universidad Nacional Aut\'{o}noma de Mexico, Ciudad Universitaria CP 04510, Mexico\label{unam}\\
\and SUPA School of Physics \& Astronomy, University of St Andrews, North Haugh, St Andrews, KY16 9SS, United Kingdom \label{standrews}\\
\and European Southern Observatory, Karl-Schwarzschild Stra\ss e 2, 85748 Garching bei M\"{u}nchen, Germany\label{eso}\\
}

   \date{Received ... ; accepted ...}

 \authorrunning{N. Kains et al.}
 \titlerunning{Mode-switching in M 68 RR Lyrae stars}
 
 \abstract{We build on our detailed analysis of time-series observations of the globular cluster M 68 to investigate the irregular pulsational behaviour of four of the RR Lyrae stars in this cluster. M 68 is one of only two globular clusters in which mode-switching of RR Lyrae stars has previously been reported, and we discuss one additional case, as well as a case of irregular behaviour, and we briefly revisit the two previously reported cases with a homogeneous analysis. We find that in 2013, V45 was pulsating in the first-overtone mode only, despite being previously reported as a double-mode (fundamental and first overtone) pulsator in 1994, and that the amplitude of the fundamental mode in V7 is increasing with time. We also suggest that V21 might not have switched pulsation modes as previously reported, although the first overtone seems to be becoming less dominant. Finally, our analysis of available archival data confirms that V33 lost a pulsation mode between 1950 and 1986.}


   \maketitle
%

\section{Introduction}\label{sec:intro}

RRL stars have been the subject of intense study for more than a century, meaning that for many of these objects, time-series photometric data are available with long baselines. This 
allows for the study of any changes in the pulsational behaviour, including period changes, Blazhko-type modulation, and mode switching. This is particularly true for RRL stars in globular 
clusters, as clusters are an ideal environment for stellar population studies. Indeed, observing campaigns to monitor globular clusters are still ongoing both from the ground 
\citep[e.g.][]{kains15, kains13b, arellano15, carretta14, kunder13b, amigo13, figuera13} and from space \citep[e.g.][]{neeley15, contreras13, montiel10}.

Only a few cases of RRL stars switching their pulsation modes have been reported in the literature.
The first cases were reported in the cluster that is the subject of this paper, M 68, by
\citeauthor{clement93} (\citeyear{clement93}, hereafter C93). Using observations by \citeauthor{vanagt59} (\citeyear{vanagt59}, hereafter VO59) from 1950 and \citeauthor{rosino54} (\citeyear{rosino54}, hereafter R54), made between 1951 and 1953, \cite{clement90} found evidence that V33 was pulsating as a RR01 star\footnote{In this paper we use the notation of \cite{alcock00} to refer to single-mode fundamental and first-overtone pulsators as RR0 and RR1, respectively, and double-mode (fundamental and first-overtone) pulsators as RR01.}. However, in observations from 1986-1991 taken and analysed by C93, V33 was found to have changed into a RR1 variable. In later years, \citeauthor{brocato94} (\citeyear{brocato94}, hereafter B94), \citeauthor{walker94} (\citeyear{walker94}, hereafter W94) and \citeauthor{kains15} (\citeyear{kains15}, hereafter K15), found that V33 continued to pulsate in the first overtone (FO) mode during observations from 1989, 1993, and 2013, respectively. C93 also found that V21 was pulsating in both the fundamental (F) and FO modes with approximately equal amplitudes over three years from 1986 to 1988, and that from 1989 to 1991 it was pulsating in the FO mode only, with no evidence for F mode pulsations.

The variable V79 in M 3 was the next case of mode-switching to be found. \cite{clement97} discovered that it had changed from RR0 in 1920-1926 \citep{larink22, muller33, greenstein35, nemec89} to RR01 in 1996, but it was then found in the 2008 observations of \cite{goranskij10} to have reverted back to single-mode pulsation in the F mode (RR0).
Also in M 3, \cite{corwin99} found that V166 had switched from F to FO pulsation over the course of one year based on observing runs in 1992 and 1993, while \cite{clementini04} found that two double-mode variables in the same cluster, V200 and V251, had changed their dominant pulsation mode over the course of the same time interval, using the data of \cite{corwin99}, but remained RR01 variables. 


Recently, \cite{soszynski14} reported on mode-switching in OGLE-BLG-RRLYR-12245, an RRL star that was first discovered from OGLE-III observations taking place between 2001 and 2006, and subsequently re-observed with the start of OGLE-IV observations. Whereas OGLE-BLG-RRLYR-12245 was a double-mode RR01 pulsator in the 2001-2006 observations, the 2010-2013 light curve shows clearly that the star was pulsating only in the F (RR0) mode. The light curve from \cite{soszynski14} also reveals a slow increase in amplitude in the 2001-2006 light curve. From the same data, \cite{soszynski14b} found two more mode-switching stars: OGLE-BLG-RRLYR-07226, which also went from RR01 to RR0, and OGLE-BLG-RRLYR-13342, which changed from RR0 to RR01. In the LMC, \cite{poleski14} found that OGLE-LMC-RRLYR-13308 had switched pulsation modes from RR01 to RR0, using OGLE and EROS-II photometry. Finally, \cite{drake14} found 6 more mode-switching RR01 stars in the Catalina Surveys of Periodic Variable Stars, in this case changing their dominant pulsation mode over just a few months, although at least one of these is in fact not a case of mode-switching \citep{soszynski14b}.

Although there is general agreement, backed up by observational evidence, that RR01 stars are associated with the region of the horizontal branch (HB) in which stars can pulsate either in the F or FO mode \citep{soszynski09}, there is no consensus on the exact causes for double-mode pulsation \citep[e.g.][]{smolec14, dziembowski93}. Studies including turbulent convection models were once thought to yield stable double-mode pulsation \citep[e.g.][]{szabo04}, but \cite{smolec08} showed that this was due to incorrect treatment of buoyant forces, and were unable to reproduce these results with the correct analysis. Furthermore, not all stars located in the either-or region are double-mode pulsators, as non-linear effect are responsible for the double-mode pulsation, meaning that the locus of RR01 stars should be narrower than the either-or region. Interestingly, however, this is not supported by the observations of K15 (see their Fig. 11, or \Fig{fig:hb} in this paper), who found that the RR01 stars in M 68 span most of the colour range of the instability strip. Indeed, \cite{arellano15} argued that the separation of the RR0 and RR1 populations (shown in \Fig{fig:hb} as a vertical dotted line), corresponds to the first overtone region's red edge (FORE), and therefore any double-mode star that is redder than this empirical border cannot be in the either-or region. Increasing the sample of detected mode-switching RRL stars is an important step towards understanding whether this behaviour is typical for stars evolving along the HB. This would also allow us to constrain the changes in internal structure that these stars are going through when moving across the instability strip, which could in turn lead to clearer explanations for other pulsational irregularities such as the Blazhko effect. Here, we report at least one additional case of mode-switching in M 68, placing this cluster on a par with M 3 with regards to the number of known cases.

M 68 (NGC~4590, C1236-264 in the IAU nomenclature; $\alpha = 12^h39^m27.98^s, \delta = -26^{\circ}44'38.6^{\prime\prime}$ at J2000.0) is one of the most metal-poor globular clusters in the Galaxy, and contains 50 known variable stars. Recently, K15 studied the variables in this cluster in detail, and used Fourier decomposition of the RR Lyrae (RRL) and period-luminosity relations for the SX Phoenicis stars to estimate the metallicity and distance of the cluster.
The 42 RRL variables in the cluster include 12 double-mode RRL (RR01) stars, which are particularly interesting, because the double-mode pulsation allows us to measure the mass of these objects \citep[e.g.][]{petersen73, popielski00, marconi15}, as well as to compare the mass-metallicity distribution of field and cluster RRL stars \citep[e.g.][]{bragaglia01}. 

In this research note, we report irregularities in the behaviour of some of the variables using archival data sets dating back over 65 years. We provide a brief summary of our 
observations in \Sec{sec:observations}, and discuss one new case of mode switching in M 68 in \Sec{sec:m68}, and irregular behaviour of one of the double-mode pulsators in \Sec{sec:v7}. In \Sec{sec:knowncases} we revisit the two previously reported cases of mode switching in M 68, and we briefly summarise our findings in \Sec{sec:conclusions}. The light curve data used in this paper are available for download via the CDS\footnote{http://cdsarc.u-strasbg.fr/viz-bin/qcat?J/A+A/578/A128}, and are linked via the relevant publications on the NASA ADS system.

\section{Observations and reductions}\label{sec:observations}

Images were obtained in $V$ (236) and $I$ (219) using CCD cameras on the 1m LCOGT telescopes at the South African Astronomical Observatory (SAAO) in Sutherland,
South Africa, as well as the Cerro Tololo Inter-American Observatory (CTIO) in Chile. The observations covered 74 days from 8th March 2013 until 20th May 2013.
The images were reduced using the {\tt DanDIA} difference image analysis software (DIA; \citealt{bramich13, bramich08}), and light curves were obtained for all stars detected
in reference images built from the best-seeing images in each filter. For further details on the observations, reductions, and calibration of the photometry, the interested reader is referred to K15.

\section{V45: A new case of mode-switching in M 68}\label{sec:m68}


\begin{figure}
  \centering
  \includegraphics[width=8cm, angle=0]{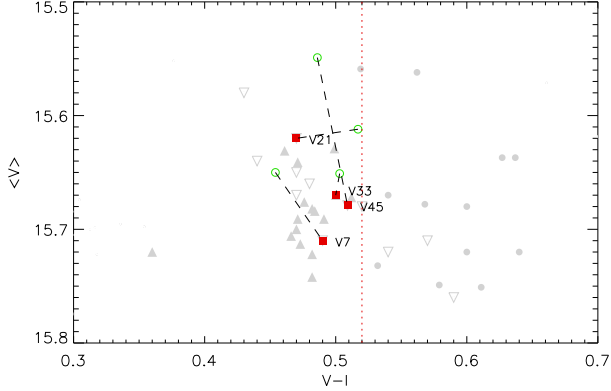}
  \caption{The HB of M 68, with RRL stars overplotted. The entire RRL population is shown in grey, with RR0 as filled circles, RR1 as filled triangles, and RR01 as inverted open triangles. The four stars discussed in this paper are shown as red filled squares (K15 observations) and open green circles (W94), with dashed lines linking their K15 and W94 positions. The vertical dotted line indicates the separation between the loci of RR0 and RR1 stars. \label{fig:hb}}
\end{figure}

\begin{figure}
  \centering
  \includegraphics[width=9cm, angle=0]{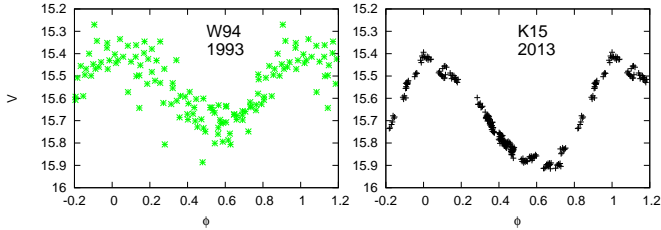}
  \caption{$V-$band light curves for V45 showing the data of W94 (left) and K15 (right), phased with the FO periods listed in \Tab{tab:v45}. \label{fig:v45}}
\end{figure}

\noindent
V45 was first detected by W94, who classified it as a double-mode RR01 star, with periods $P_0=0.5239$ d and $P_1=0.39087$ d, giving a period ratio of 0.746. However, K15 found no evidence of a second pulsation mode in their recent data (\Tab{tab:v45}). Comparing the light curves of W94 with those of K15 (\Fig{fig:v45}) suggests indeed that the double-mode pulsation that was present in the 1993 observations essentially disappeared between 1993 and 2013, with the most recent light curve looking like that of an RR1 star with period of $0.3908187$ d. The position of V45 on the HB was significantly different in the data of K15 compared to the observations of W94. Its location near several bright stars (see Fig. 10 of K15) means that it is likely that W94 over-estimated the brightness of V45 because of blending, which had little effect on the photometry of K15 thanks to their use of difference image analysis \citep[e.g.][]{kains12b, bramich12}. We note here that K15 underestimated the average brightness of V45 in $V$ by 0.04 mag, and therefore its $V-I$ colour. Here we revise this value to find $\langle V\rangle=15.68 \pm 0.01$ mag, and $V-I = 0.51 \pm 0.01$ mag, placing it among the RR1 population.

\begin{table*}
\begin{center}
  \begin{tabular}{cccccccccccc}

   \hline
    Data		&Observed	&$A_B$ 	&$A_V$	&$A_I$ 	&$P_0$	&$P_1$	&$k_B$	&$k_V$	&$k_I$ &Type \\
 		&(year)	&(mag)	&(mag) 	&(mag) 	&(d)	&(d)	 \\
  \hline  
    W94		&1993		&0.74	&$\sim$0.47	&$\sim$0.45	&0.5076(2)	&0.38450(4)	&4.7(9)	&3.0(6)	&5(2)	&RR01\\
    K15		&2013		&$-$		&0.52		&0.30		&$-$			&0.39082(1)	&$-$		&$-$		&$-$		&RR1\\
\hline
  \end{tabular}
  \caption{Same as \Tab{tab:v7}, but for V45. \label{tab:v45}}
  \end{center}
\end{table*}

Performing a Fourier analysis of the data of W94, we found an amplitude for the F mode pulsation of 0.045 mag in $V$ and 0.022 mag in $I$. The residual Fourier noise of the K15 light curve, after prewhitening the FO pulsation, is 0.008 mag in $V$, and 0.006 mag in $I$ (see \Tab{tab:fouriernoise}), meaning that if present in 2013, the F-mode pulsation would have been detected in both filters, unless the F mode had dropped below detection limits.

\section{Amplitude ratio variations in V7}\label{sec:v7}

\begin{table*}
\begin{center}
  \begin{tabular}{cccccccccccccc}

     \hline
    Data		&Observed	&$A_{\rm p}$	&$A_B$ 	&$A_V$	&$A_I$ 	&$P_0$	&$P_1$	&$k_{\rm p}$	&$k_B$	&$k_V$	&$k_I$ &Type \\
    		&(year)	&(mag)	&(mag) 	&(mag)	&(mag) 	&(d)	&(d)	&	&	&	& 	& \\
  \hline  
    VO59		&1950		&0.33	&$-$		&$-$		&$-$		&0.5174(1)	&0.3881(6)	&3.0(5)	&$-$		&$-$		&$-$	&RR01	\\
    R54		&1951-1953	&0.36	&$-$		&$-$		&$-$		&0.54118(4)	&0.387951(5)	&5.1(9)	&$-$		&$-$		&$-$	&RR01?   \\
    C93		&1986-1991	&$-$		&0.87	&$-$		&$-$		&0.519742(8)	&0.387958(2)	&$-$		&3.0(3)	&$-$		&$-$	&RR01	\\
    W94		&1993		&$-$		&0.87	&0.63	&0.53	&0.51929(8)	&0.38801(2)	&$-$		&3.0(3)	&2.5(2)	&1.4(1)	&RR01	\\
    K15		&2013		&$-$		&$-$		&0.62	&0.41	&0.52028(8)	&0.38812(2)	&$-$		&$-$		&1.8(1)	&1.7(1)	&RR01	\\
\hline
  \end{tabular}
  \caption{Light curve amplitudes (peak-to-peak for the filtered light curves, Fourier amplitudes for the photographic data) of the phased light curves and periods of the double-mode variable V7 in the different available data sets. All periods have been recalculated with the software {\tt Period04}, and amplitudes were also re-measured from the light curves. The ratio of the FO to F amplitudes, $k$, is also given for each filter, and p denotes photographic data. We do not include B94 data because of poor phase coverage. The number is parentheses next to each value is the uncertainty on the last decimal point. \label{tab:v7}}
  \end{center}
\end{table*}

\begin{figure}
  \centering
  \includegraphics[width=4.8cm, angle=0]{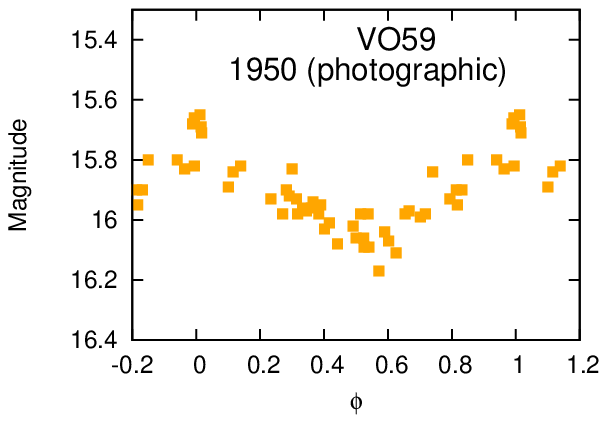}
  \includegraphics[width=9cm, angle=0]{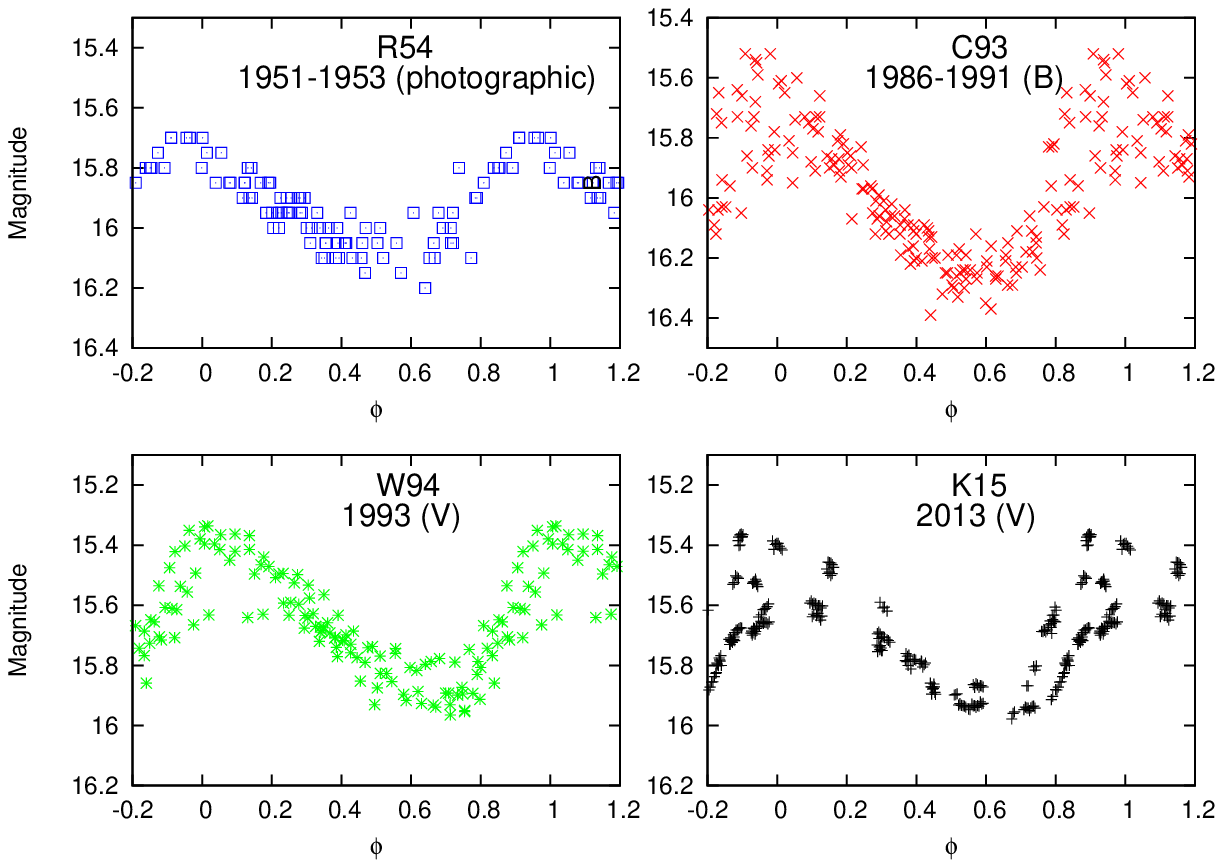}
  \caption{Phased light curves (with the FO periods given in \Tab{tab:v7}) for V7 showing the data of VO59 (unfiltered photographic plates, top), R54 (unfiltered photographic plates, middle left), C93 ($B$-band, middle right), W94 ($V$, bottom left), and K15 ($V$, bottom right). The data of B94 are not plotted because they have poor phase coverage. \label{fig:v7_all}}
\end{figure}

\noindent
V7 was discovered by \cite{shapley19}, and was first identified as a double-mode pulsator by C93. K15 calculated F and FO periods of $P_0=0.520186$ d and $P_1=0.3879608$ d, with a period ratio of 0.746, typical of RR01 pulsators \citep[e.g][]{poleski14}, and similar to values found by C93 and W94. Its position on the HB, along with the other three stars discussed in this paper, is shown in \Fig{fig:hb}.

Comparing our data with archival data, we noticed that the photographic plate light curve amplitudes of VO59 and R54 are 0.33 and 0.36 mag, respectively, whereas 
the amplitude in $B$ found by C93 was 0.87 mag, and in $V$, K15 found an amplitude of 0.62 mag, similar to the $V$ amplitude of 0.63 mag found by W94. This is surprising, because amplitudes measured from photographic plates should be larger than in the redder $V$ filter. This suggests that the amplitude of one of the pulsation modes has grown in the time between the observations of VO59 and R54, and those of C93, W94, and K15. The photographic light curves taken in the 1950s do not show as large scatter due to double-mode pulsation as the more recent data sets, and R54 classified V7 as a RR1 star. 

The results of a homogeneous re-analysis of all available light curves for V7 with the period analysis software {\tt Period04} \citep{lenz05} are given in \Tab{tab:v7}. The ratio of the FO to F mode amplitudes, $k$, is given for each data set and each filter. This ratio decreases with time, supporting our hypothesis that the F mode is growing over time. By comparing the amplitudes of the two modes, we note, however, that the amplitude of the F mode for the case of K15 is $\sim$0.4 mag in $V$, which is too small for the F mode period when compared to the locus of RR0 stars on the period-amplitude diagram of M 68 (Fig. 17 of K15). This indicates that the amplitudes of each mode in double-mode pulsators are smaller than their single-mode counterparts, which is expected if non-resonant mode coupling takes place \citep[e.g.][]{smolec14}, as is expected to be the case in a majority of double-mode stars. In any case, even if this indicates that the F mode is weak, it is still becoming stronger relative to the FO mode.

In $I$, the ratio increased from W94 to K15, but we believe $k_I$ to be underestimated in the data of W94 due to a poor light curve. Although we do detect evidence for F mode pulsation in the R54 data, we find a period for this mode of 0.541177 d, which is much longer than the fundamental period derived from any of the other data sets. This could indicate that a mode switch was about to happen, or had recently happened, in a similar manner to V79 in M 3 \citep{goranskij10}, but we lack the observations to verify this possibility. Furthermore, the F mode detection is weak, with a ratio $k$ of 5.16, leaving open the possibility that V7 might have been pulsating as a RR1 star in the observations of R54.

We note that V7 has the shortest periods (both $P_0$ and $P_1$) of all RR01 stars in M 68 (K15). The periods derived for the pulsation modes of V7 also seem to vary significantly with time, particularly the F mode period.

\begin{table*}
\begin{center}
  \begin{tabular}{cc|cc|cc|cc|cc}

\hline
    			&		&V7		&		&V21 	& 		&V33 	& 		&V45 	& \\
  \hline
    Data		&Filter	&FO		&F		&FO 		&F 		&FO 		&F 		&FO 		&F \\
  \hline  
    VO59		&$p$		&0.011	&0.012	&0.038	&$-$		&$-$		&$-$		&$-$		&$-$\\
    R54		&$p$		&0.013	&0.012	&$-$		&$-$		&$-$		&$-$		&$-$		&$-$\\
    C93A		&$B$	&$-$		&$-$		&0.037	&0.033	&$-$		&$-$		&$-$		&$-$\\
    C93B		&$B$	&$-$		&$-$		&0.049	&$-$		&$-$		&$-$		&$-$		&$-$\\
    C93		&$B$	&0.012	&0.011	&0.040	&0.042	&0.012	&$-$		&$-$		&$-$\\
    W94		&$B$	&0.014	&0.012	&0.018	&0.018	&0.009	&$-$		&0.025	&0.024\\
    W94		&$V$	&0.010	&0.009	&0.014	&0.014	&0.007	&$-$		&0.017	&0.014\\
    W94		&$I$		&0.013	&0.012	&0.008	&0.009	&0.005	&$-$		&0.016	&0.016\\
    K15		&$V$	&0.015	&0.015	&0.016	&0.016	&0.009	&$-$		&0.008	&$-$\\
    K15		&$I$		&0.010	&0.010	&0.010	&0.010	&0.004	&$-$		&0.006	&$-$\\
\hline
  \end{tabular}
  \caption{Fourier noise after prewhitening, in magnitudes, at the F and FO frequencies for the four variables discussed in this paper, for each filter and each data set. \label{tab:fouriernoise}}
  \end{center}
\end{table*}

\section{Previously reported cases}\label{sec:knowncases} 

\subsection{V21}


\begin{figure}
  \centering
  \includegraphics[width=9cm, angle=0]{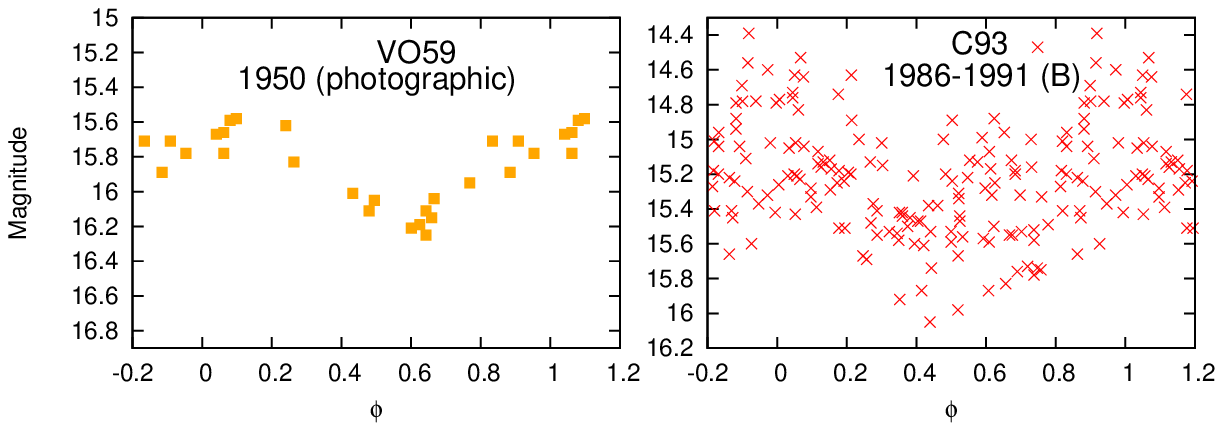}
  \includegraphics[width=9cm, angle=0]{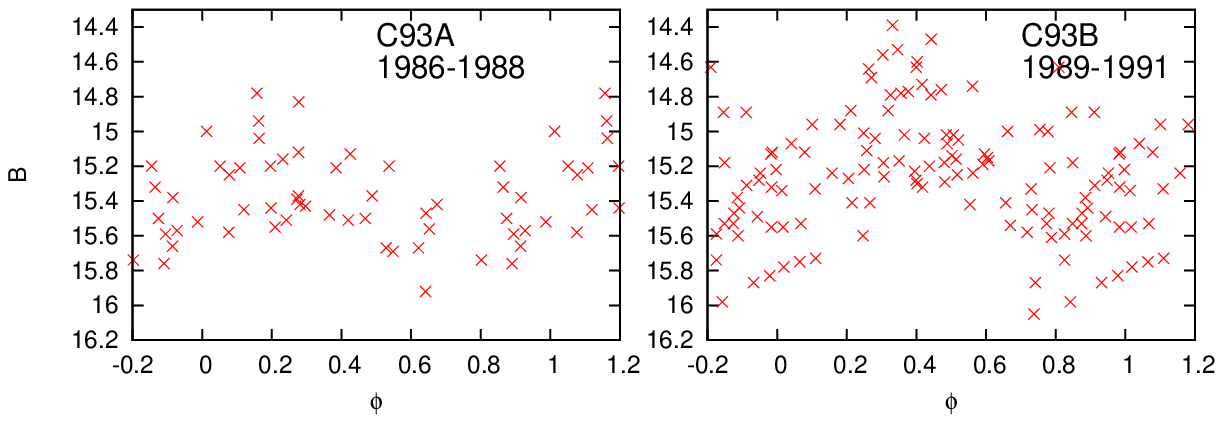}
  \includegraphics[width=9cm, angle=0]{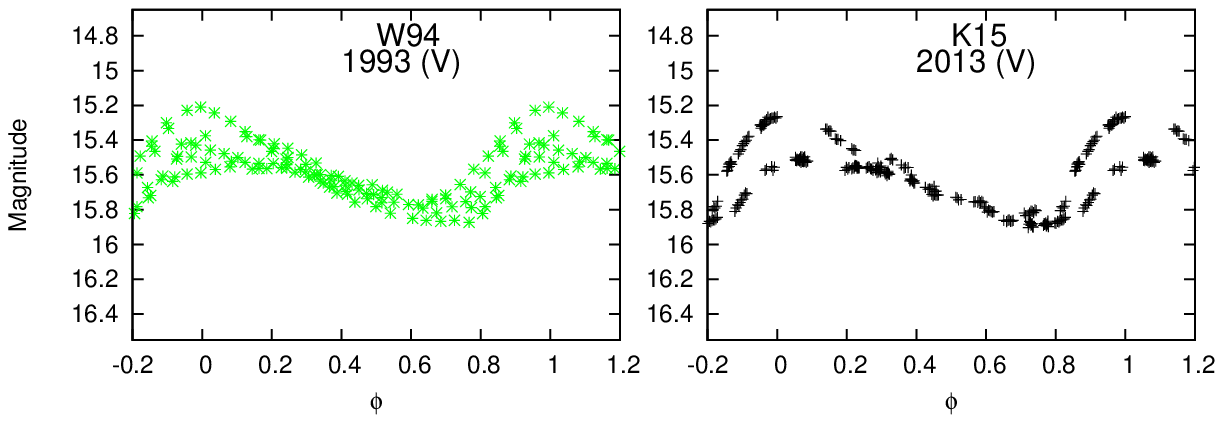}
  \caption{Light curves for V21 showing the data of VO59 (unfiltered photographic, top left), C93 ($B$, top right, C93A and C93B are shown separately in the middle panels), W94 ($V$, bottom left), and K15 ($V$, bottom right). The light curves are phased with the FO periods listed in \Tab{tab:v21}. \label{fig:v21_all}}
\end{figure}

\noindent
As mentioned in \Sec{sec:intro}, V21 was found by C93 to have switched from being an RR01 pulsator between 1986 and 1988 to an RR1 star between 1989 and 1991, meaning that the transition must have been abrupt. However, both W94 and K15 found that V21 was once again a clear double-mode RR01 pulsator, meaning that another transition must then have occurred between 1991 and 1993. We calculated Fourier fits using {\tt Period04} in order to determine the amplitude of each mode in the data sets of C93, W94, and K15. We also considered the 1986-1988 and 1989-1991 parts of the C93 light curve separately, to which we refer as C93A and C93B respectively. 

Although our analysis does confirm that two modes are detected in C93A, and only one in C93B (\Tab{tab:v21}), we suggest that the poor quality of the C93 light curve (\Fig{fig:v21_all}) does not actually allow a strong conclusion to be made as to the pulsational behaviour of V21. The evidence for an abrupt mode switch from RR01 to RR1 and then back to RR01 over the course of 5 years is therefore weak. In addition to this, analysis of the C93A data suggests that the F mode was dominant over 1986-1988, which is unusual for an RR01 pulsator \citep{smolec15}, and the ratio of periods of the detected modes is 0.775, significantly higher than the canonical value of $\sim$0.746.

However, there is evidence of interesting behaviour in the V21 light curves. We re-analysed the light curves of VO59 and were unable to find signs of the F mode pulsation, although the light curve is likely too sparse to draw conclusions as to the mode content of V21 in 1950. We also found that the amplitude ratio $k$ decreased in both $V$ and $I$ between the W94 and K15 data. Since very similar light curve amplitudes were found in those two studies in both $V$ and $I$, along with very similar periods (\Tab{tab:v21}), this change is likely to be real, suggesting that the relative amplitude of the FO to the F mode has decreased between 1993 and 2013.\\

\begin{table*}
\begin{center}
  \begin{tabular}{cccccccccccccc}

     \hline
    Data		&Observed	&$A_{\rm p}$	&$A_B$ 	&$A_V$	&$A_I$ 	&$P_0$	&$P_1$	&$k_{\rm p}$	&$k_B$	&$k_V$	&$k_I$ &Type \\
  		&(year)	&(mag)	&(mag) 	&(mag)	&(mag) 	&(d)	&(d)	&	&	&	& 	& \\
  \hline  
    VO59		&1950		&0.7		&$-$		&$-$		&$-$		&$-$			&0.3821(2)	&$-$		&$-$		&$-$		&$-$	&RR1	\\
    C93A		&1986-1988	&$-$		&1.14	&$-$		&$-$		&0.52566(5)	&0.40716(3)	&$-$		&0.8(2)	&$-$		&$-$	&RR01?	\\
    C93B		&1989-1991	&$-$		&1.66	&$-$		&$-$		&$-$			&0.40374(1)	&$-$		&$-$		&$-$		&$-$	&RR1	\\
    C93		&1986-1991	&$-$		&1.66	&$-$		&$-$		&0.52783(2)	&0.40036(1)	&$-$		&1.7(4)	&$-$		&$-$	&RR01	\\
    W94		&1993		&$-$		&0.85	&0.66	&0.41	&0.54577(9)	&0.40717(2)	&$-$		&2.0(1)	&2.0(1)	&1.9(2) &RR01	\\
    K15		&2013		&$-$		&$-$		&0.64	&0.42	&0.54575(7)	&0.40711(2)	&$-$		&$-$		&1.6(1)	&1.5(1) &RR01	\\
\hline
  \end{tabular}
  \caption{Same as \Tab{tab:v7}, but for V21.  \label{tab:v21}}
  \end{center}
\end{table*}

\subsection{V33}

We re-analysed the data of VO59 for this star using {\tt Period04}, and confirm that there are signs of double-mode pulsation in that data set, as first suggested by \cite{clement90}. We also find a FO period for the VO59 data that is significantly shorter than those subsequently found by C93, W94 and K15, when the star was a RR1 pulsator (\Tab{tab:v33}). Therefore, we confirm that V33 switched from being a RR01 to RR1 star between 1950 and 1986. We also note that the FO period is becoming longer with time.

\begin{figure}

  \includegraphics[width=9cm, angle=0]{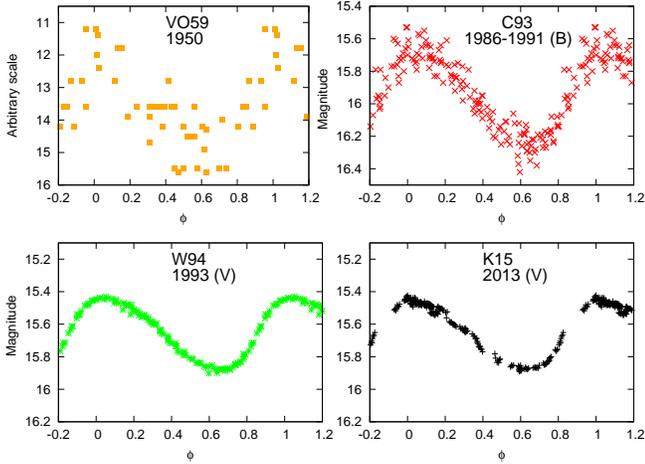}
  \caption{Light curves for V33 showing the data of VO59 (unfiltered photographic, top left), C93 ($B$, top right), W94 ($V$, bottom left), and K15 ($V$, bottom right). The light curves are phased with the FO periods listed in \Tab{tab:v33}. \label{fig:v33_all}}
\end{figure}

\begin{table*}
\begin{center}
  \begin{tabular}{ccccccccccccc}

     \hline
    Data		&Observed	&$A_{\rm p}$	&$A_B$ 	&$A_V$	&$A_I$ 	&$P_0$	&$P_1$	&$k_{\rm p}$	&Type \\
  		&(year)	&(mag)	&(mag) 	&(mag)	&(mag) 	&(d)	&(d)	&	&	&	& 	& \\
  \hline  
    VO59		&1950		&$-$		&$-$		&$-$		&$-$		&0.5088(9)	&0.3852(2)	&2.7(8)	&RR01	\\
    C93		&1986-1991	&$-$		&0.89	&$-$		&$-$		&$-$			&0.390572(1)	&$-$		&RR1	\\ 
    W94		&1993		&$-$		&0.59	&0.47	&0.32	&$-$			&0.39059(1)	&$-$		&RR1	\\ 
    K15		&2013		&$-$		&$-$		&0.47	&0.23	&$-$			&0.39066(1)	&$-$		&RR1	\\ 
\hline
  \end{tabular}
  \caption{Same as \Tab{tab:v7}, but for V33. No amplitude is given for the VO59 data as they are not on a magnitude scale. \label{tab:v33}}
  \end{center}
\end{table*}

\section{Conclusions}\label{sec:conclusions}

Further comparative analysis of the data of K15 with archival data revealed interesting behaviour in several RRL stars. The F mode in V45 has either disappeared or become very weak, meaning that it now pulsates like a RR1 star, with the switch having occurred between 1993 and 2013. We find that the F mode in V7 is growing, and that a mode-switching event might have taken place in the 1950s, associated with a large change in the F pulsation period, but the evidence is too weak to make strong conclusions on this for V7. The F mode of V21 has grown with time, although again the claim of mode switching is weak, as it relies solely on a light curve that suffers from large scatter. We also find evidence to support the claim of C93 that V33 switched modes from RR01 to RR1 between 1950 and 1986. Interestingly, the double-mode stars in M 68 are distributed over a large part of the instability strip, which does not support the current understanding of double-mode pulsation.
   
\section*{Acknowledgements}

We thank Christine Clement for sharing with us her files containing the light curves of VO59, and the anonymous referee for their time and constructive feedback. DMB acknowledges NPRP grant \# X-019-1-006 from the Qatar National Research Fund (a member of Qatar Foundation). AAF acknowledges the support of DGAPA-UNAM through project IN106615-17.

\bibliographystyle{aa}
\bibliography{../thesisbib}

\begin{thebibliography}{45}
\expandafter\ifx\csname natexlab\endcsname\relax\def\natexlab#1{#1}\fi

\bibitem[{{Alcock} {et~al.}(2000){Alcock}, {Allsman}, {Alves}, {Axelrod},
  {Becker}, {Bennett}, {Clement}, {Cook}, {Drake}, {Freeman}, {Geha}, {Griest},
  {Kov{\'a}cs}, {Kurtz}, {Lehner}, {Marshall}, {Minniti}, {Nelson}, {Peterson},
  {Popowski}, {Pratt}, {Quinn}, {Rodgers}, {Rowe}, {Stubbs}, {Sutherland},
  {Tomaney}, {Vandehei}, \& {Welch}}]{alcock00}
{Alcock}, C., {Allsman}, R., {Alves}, D.~R., {et~al.} 2000, \apj, 542, 257

\bibitem[{{Amigo} {et~al.}(2013){Amigo}, {Stetson}, {Catelan}, {Zoccali}, \&
  {Smith}}]{amigo13}
{Amigo}, P., {Stetson}, P.~B., {Catelan}, M., {Zoccali}, M., \& {Smith}, H.~A.
  2013, \aj, 146, 130

\bibitem[{{Arellano Ferro} {et~al.}(2015){Arellano Ferro}, {Mancera Pi{\~n}a},
  {Bramich}, {Giridhar}, {Ahumada}, {Kains}, \& {Kuppuswamy}}]{arellano15}
{Arellano Ferro}, A., {Mancera Pi{\~n}a}, P.~E., {Bramich}, D.~M., {et~al.}
  2015, \mnras, 452, 727

\bibitem[{{Bragaglia} {et~al.}(2001){Bragaglia}, {Gratton}, {Carretta},
  {Clementini}, {Di Fabrizio}, \& {Marconi}}]{bragaglia01}
{Bragaglia}, A., {Gratton}, R.~G., {Carretta}, E., {et~al.} 2001, \aj, 122, 207

\bibitem[{{Bramich}(2008)}]{bramich08}
{Bramich}, D.~M. 2008, \mnras, 386, L77

\bibitem[{{Bramich} {et~al.}(2012){Bramich}, {Arellano Ferro}, {Figuera
  Jaimes}, \& {Giridhar}}]{bramich12}
{Bramich}, D.~M., {Arellano Ferro}, A., {Figuera Jaimes}, R., \& {Giridhar}, S.
  2012, \mnras, 424, 2722

\bibitem[{{Bramich} {et~al.}(2013){Bramich}, {Horne}, {Albrow}, {Tsapras},
  {Snodgrass}, {Street}, {Hundertmark}, {Kains}, {Arellano}, {Figuera}, \&
  {Giridhar}}]{bramich13}
{Bramich}, D.~M., {Horne}, K., {Albrow}, M.~D., {et~al.} 2013, \mnras, 428,
  2275

\bibitem[{{Brocato} {et~al.}(1994){Brocato}, {Castellani}, \&
  {Ripepi}}]{brocato94}
{Brocato}, E., {Castellani}, V., \& {Ripepi}, V. 1994, \aj, 107, 622

\bibitem[{{Carretta} {et~al.}(2014){Carretta}, {Bragaglia}, {Gratton},
  {D'Orazi}, {Lucatello}, {Momany}, {Sollima}, {Bellazzini}, {Catanzaro}, \&
  {Leone}}]{carretta14}
{Carretta}, E., {Bragaglia}, A., {Gratton}, R.~G., {et~al.} 2014, \aap, 564,
  A60

\bibitem[{{Clement}(1990)}]{clement90}
{Clement}, C.~M. 1990, \aj, 99, 240

\bibitem[{{Clement} {et~al.}(1993){Clement}, {Ferance}, \& {Simon}}]{clement93}
{Clement}, C.~M., {Ferance}, S., \& {Simon}, N.~R. 1993, \apj, 412, 183

\bibitem[{{Clement} \& {Shelton}(1997)}]{clement97}
{Clement}, C.~M. \& {Shelton}, I. 1997, \aj, 113, 1711

\bibitem[{{Clementini} {et~al.}(2004){Clementini}, {Corwin}, {Carney}, \&
  {Sumerel}}]{clementini04}
{Clementini}, G., {Corwin}, T.~M., {Carney}, B.~W., \& {Sumerel}, A.~N. 2004,
  \aj, 127, 938

\bibitem[{{Contreras Ramos} {et~al.}(2013){Contreras Ramos}, {Clementini},
  {Federici}, {Fiorentino}, {Cacciari}, {Catelan}, {Smith}, {Fusi Pecci},
  {Marconi}, {Pritzl}, \& {Kinemuchi}}]{contreras13}
{Contreras Ramos}, R., {Clementini}, G., {Federici}, L., {et~al.} 2013, \apj,
  765, 71

\bibitem[{{Corwin} {et~al.}(1999){Corwin}, {Carney}, \& {Allen}}]{corwin99}
{Corwin}, T.~M., {Carney}, B.~W., \& {Allen}, D.~M. 1999, \aj, 117, 1332

\bibitem[{{Drake} {et~al.}(2014){Drake}, {Graham}, {Djorgovski}, {Catelan},
  {Mahabal}, {Torrealba}, {Garc{\'{\i}}a-{\'A}lvarez}, {Donalek}, {Prieto},
  {Williams}, {Larson}, {Christen sen}, {Belokurov}, {Koposov}, {Beshore},
  {Boattini}, {Gibbs}, {Hill}, {Kowalski}, {Johnson}, \& {Shelly}}]{drake14}
{Drake}, A.~J., {Graham}, M.~J., {Djorgovski}, S.~G., {et~al.} 2014, \apjs,
  213, 9

\bibitem[{{Dziembowski}(1993)}]{dziembowski93}
{Dziembowski}, W.~A. 1993, in Astronomical Society of the Pacific Conference
  Series, Vol.~40, IAU Colloq. 137: Inside the Stars, ed. W.~W. {Weiss} \&
  A.~{Baglin}, 521--534

\bibitem[{{Figuera Jaimes} {et~al.}(2013){Figuera Jaimes}, {Arellano Ferro},
  {Bramich}, {Giridhar}, \& {Kuppuswamy}}]{figuera13}
{Figuera Jaimes}, R., {Arellano Ferro}, A., {Bramich}, D.~M., {Giridhar}, S.,
  \& {Kuppuswamy}, K. 2013, \aap, 556, A20

\bibitem[{{Goranskij} {et~al.}(2010){Goranskij}, {Clement}, \&
  {Thompson}}]{goranskij10}
{Goranskij}, V., {Clement}, C.~M., \& {Thompson}, M. 2010, in Variable Stars,
  the Galactic halo and Galaxy Formation, ed. C.~{Sterken}, N.~{Samus}, \&
  L.~{Szabados}, 115

\bibitem[{{Greenstein}(1935)}]{greenstein35}
{Greenstein}, J.~L. 1935, Astronomische Nachrichten, 257, 301

\bibitem[{{Kains} {et~al.}(2015){Kains}, {Arellano Ferro}, {Figuera Jaimes},
  {Bramich}, {Skottfelt}, {J{\o}rgensen}, {Tsapras}, {Street}, {Browne},
  {Dominik}, {Horne}, {Hundertmark}, {Ipatov}, {Snodgrass}, {Steele},
  {Alsubai}, {Bozza}, {Calchi Novati}, {Ciceri}, {D'Ago}, {Galianni}, {Gu},
  {Harps{\o}e}, {Hinse}, {Juncher}, {Korhonen}, {Mancini}, {Popovas}, {Rabus},
  {Rahvar}, {Southworth}, {Surdej}, {Vilela}, {Wang}, \& {Wertz}}]{kains15}
{Kains}, N., {Arellano Ferro}, A., {Figuera Jaimes}, R., {et~al.} 2015, \aap,
  578, A128

\bibitem[{{Kains} {et~al.}(2013){Kains}, {Bramich}, {Arellano Ferro}, {Figuera
  Jaimes}, {J{\o}rgensen}, {Giridhar}, {Penny}, {Alsubai}, {Andersen}, {Bozza},
  {Browne}, {Burgdorf}, {Calchi Novati}, {Damerdji}, {Diehl}, {Dodds},
  {Dominik}, {Elyiv}, {Fang}, {Giannini}, {Gu}, {Hardis}, {Harps{\o}e},
  {Hinse}, {Hornstrup}, {Hundertmark}, {Jessen-Hansen}, {Juncher}, {Kerins},
  {Kjeldsen}, {Korhonen}, {Liebig}, {Lund}, {Lundkvist}, {Mancini}, {Martin},
  {Mathiasen}, {Rabus}, {Rahvar}, {Ricci}, {Sahu}, {Scarpetta}, {Skottfelt},
  {Snodgrass}, {Southworth}, {Surdej}, {Tregloan-Reed}, {Vilela}, {Wertz}, \&
  {Williams}}]{kains13b}
{Kains}, N., {Bramich}, D.~M., {Arellano Ferro}, A., {et~al.} 2013, \aap, 555,
  A36

\bibitem[{{Kains} {et~al.}(2012){Kains}, {Bramich}, {Figuera Jaimes}, {Arellano
  Ferro}, {Giridhar}, \& {Kuppuswamy}}]{kains12b}
{Kains}, N., {Bramich}, D.~M., {Figuera Jaimes}, R., {et~al.} 2012, \aap, 548,
  A92

\bibitem[{{Kunder} {et~al.}(2013){Kunder}, {Stetson}, {Cassisi}, {Layden},
  {Bono}, {Catelan}, {Walker}, {Paredes Alvarez}, {Clem}, {Matsunaga},
  {Salaris}, {Lee}, \& {Chaboyer}}]{kunder13b}
{Kunder}, A., {Stetson}, P.~B., {Cassisi}, S., {et~al.} 2013, \aj, 146, 119

\bibitem[{{Larink}(1922)}]{larink22}
{Larink}, J. 1922, Astronomische Abhandlungen der Hamburger Sternwarte, 2, 1

\bibitem[{{Lenz} \& {Breger}(2005)}]{lenz05}
{Lenz}, P. \& {Breger}, M. 2005, Communications in Asteroseismology, 146, 53

\bibitem[{{Marconi} {et~al.}(2015){Marconi}, {Coppola}, {Bono}, {Braga},
  {Pietrinferni}, {Buonanno}, {Castellani}, {Musella}, {Ripepi}, \&
  {Stellingwerf}}]{marconi15}
{Marconi}, M., {Coppola}, G., {Bono}, G., {et~al.} 2015, ArXiv e-prints

\bibitem[{{Montiel} \& {Mighell}(2010)}]{montiel10}
{Montiel}, E.~J. \& {Mighell}, K.~J. 2010, \aj, 140, 1500

\bibitem[{{Muller}(1933)}]{muller33}
{Muller}, T. 1933, Veroeffentlichungen der Universitaetssternwarte zu
  Berlin-Babelsberg, 1

\bibitem[{{Neeley} {et~al.}(2015){Neeley}, {Marengo}, {Bono}, {Braga},
  {Dall'Ora}, {Stetson}, {Ferraro}, {Freedman}, {Iannicola}, {Madore},
  {Matsunaga}, {Monson}, {Persson}, {Scowcroft}, \& {Seibert}}]{neeley15}
{Neeley}, J.~R., {Marengo}, M., {Bono}, G., {et~al.} 2015, ArXiv e-prints

\bibitem[{{Nemec} \& {Clement}(1989)}]{nemec89}
{Nemec}, J.~M. \& {Clement}, C.~M. 1989, \aj, 98, 860

\bibitem[{{Petersen}(1973)}]{petersen73}
{Petersen}, J.~O. 1973, \aap, 27, 89

\bibitem[{{Poleski}(2014)}]{poleski14}
{Poleski}, R. 2014, \pasp, 126, 509

\bibitem[{{Popielski} {et~al.}(2000){Popielski}, {Dziembowski}, \&
  {Cassisi}}]{popielski00}
{Popielski}, B.~L., {Dziembowski}, W.~A., \& {Cassisi}, S. 2000, \actaa, 50,
  491

\bibitem[{{Rosino} \& {Pietra}(1954)}]{rosino54}
{Rosino}, L. \& {Pietra}, S. 1954, \memsai, 25, 227

\bibitem[{{Shapley}(1919)}]{shapley19}
{Shapley}, H. 1919, \pasp, 31, 226

\bibitem[{{Smolec}(2014)}]{smolec14}
{Smolec}, R. 2014, in IAU Symposium, Vol. 301, IAU Symposium, ed. J.~A.
  {Guzik}, W.~J. {Chaplin}, G.~{Handler}, \& A.~{Pigulski}, 265--272

\bibitem[{{Smolec} \& {Moskalik}(2008)}]{smolec08}
{Smolec}, R. \& {Moskalik}, P. 2008, \actaa, 58, 233

\bibitem[{{Smolec} {et~al.}(2015){Smolec}, {Soszy{\'n}ski}, {Udalski},
  {Szyma{\'n}ski}, {Pietrukowicz}, {Skowron}, {Koz{\l}owski}, {Poleski},
  {Skowron}, {Pietrzy{\'n}ski}, {Wyrzykowski}, {Ulaczyk}, \&
  {Mr{\'o}z}}]{smolec15}
{Smolec}, R., {Soszy{\'n}ski}, I., {Udalski}, A., {et~al.} 2015, \mnras, 447,
  3756

\bibitem[{{Soszy{\'n}ski} {et~al.}(2014{\natexlab{a}}){Soszy{\'n}ski},
  {Dziembowski}, {Udalski}, {Szyma{\'n}ski}, {Kubiak}, {Pietrzy{\'n}ski},
  {Wyrzykowski}, {Ulaczyk}, {Poleski}, {Koz{\l}owski}, {Pietrukowicz},
  {Skowron}, \& {Mr{\'o}z}}]{soszynski14}
{Soszy{\'n}ski}, I., {Dziembowski}, W.~A., {Udalski}, A., {et~al.}
  2014{\natexlab{a}}, \actaa, 64, 1

\bibitem[{{Soszy{\'n}ski} {et~al.}(2009){Soszy{\'n}ski}, {Udalski},
  {Szyma{\'n}ski}, {Kubiak}, {Pietrzy{\'n}ski}, {Wyrzykowski}, {Szewczyk},
  {Ulaczyk}, \& {Poleski}}]{soszynski09}
{Soszy{\'n}ski}, I., {Udalski}, A., {Szyma{\'n}ski}, M.~K., {et~al.} 2009,
  \actaa, 59, 1

\bibitem[{{Soszy{\'n}ski} {et~al.}(2014{\natexlab{b}}){Soszy{\'n}ski},
  {Udalski}, {Szyma{\'n}ski}, {Pietrukowicz}, {Mr{\'o}z}, {Skowron},
  {Koz{\l}owski}, {Poleski}, {Skowron}, {Pietrzy{\'n}ski}, {Wyrzykowski},
  {Ulaczyk}, \& {Kubiak}}]{soszynski14b}
{Soszy{\'n}ski}, I., {Udalski}, A., {Szyma{\'n}ski}, M.~K., {et~al.}
  2014{\natexlab{b}}, \actaa, 64, 177

\bibitem[{{Szab{\'o}} {et~al.}(2004){Szab{\'o}}, {Koll{\'a}th}, \&
  {Buchler}}]{szabo04}
{Szab{\'o}}, R., {Koll{\'a}th}, Z., \& {Buchler}, J.~R. 2004, \aap, 425, 627

\bibitem[{{van Agt} \& {Oosterhoff}(1959)}]{vanagt59}
{van Agt}, S.~L.~T.~J. \& {Oosterhoff}, P.~T. 1959, Annalen van de Sterrewacht
  te Leiden, 21, 253

\bibitem[{{Walker}(1994)}]{walker94}
{Walker}, A.~R. 1994, \aj, 108, 555

\end{thebibliography}

\label{lastpage}

\end{document}